\documentclass[a4paper,fleqn]{cas-dc}
\usepackage[authoryear,round]{natbib}
\usepackage{hyperref}
\usepackage{makecell}
\usepackage{color,graphicx}
\usepackage{float}
\usepackage{multirow,amsmath,amsfonts,amssymb,graphics,psfrag,theorem,calc,url,bm}
\usepackage{stfloats}
\usepackage{algorithm,algorithmic}
\usepackage{diagbox}
\usepackage{hhline}
\usepackage{subfigure}
\usepackage{pifont}
\usepackage{booktabs}
\usepackage{arydshln}
\usepackage{orcidlink}

\usepackage{mathtools}

\makeatletter
\def\multilimits@{\bgroup
	\Let@
	\restore@math@cr
	\default@tag
	\baselineskip\fontdimen10 \scriptfont\tw@
	\advance\baselineskip\fontdimen12 \scriptfont\tw@
	\lineskip\thr@@\fontdimen8 \scriptfont\thr@@
	\lineskiplimit\lineskip
	\vbox\bgroup\ialign\bgroup\hfil$\m@th\scriptstyle{##}$\hfil\crcr}
\def\Sb{_\multilimits@}
\def\endSb{\crcr\egroup\egroup\egroup}
\makeatother

	{\end{list}}

\newlength{\twidth}
\definecolor{orange}{RGB}{255,107,0}

\theorembodyfont{\rmfamily}

{\begin{list}{}{
			\renewcommand{\makelabel}[1]{\mbox{\textnormal{##1}}\hfil}%
			\settowidth{\labelwidth}{\mbox{\textnormal{#1}}}%
			\setlength{\leftmargin}{\labelwidth+\labelsep}}}%
	{\end{list}}

\newcommand\bE{\ensuremath{{\bm E}}}

\newcommand\bI{\ensuremath{{\bm I}}}
\newcommand\bQ{\ensuremath{{\bm Q}}}

\newcommand\bs{\ensuremath{{\bm s}}}
\newcommand\bu{\ensuremath{{\bm u}}}
\newcommand\bx{\ensuremath{{\bm x}}}
\newcommand\bz{\ensuremath{{\bm z}}}
\newcommand\bpsi{\ensuremath{{\bm \psi}}}

\newcommand{\draftincludegraphics}[3][]{%
  \IfFileExists{#2}{\includegraphics[#1]{#2}}{%
    \fbox{\parbox[c][12mm][c]{\dimexpr#3-2\fboxsep-2\fboxrule\relax}{\centering\scriptsize Missing\\image}}}}
\newproof{proof}{Proof}
 
\ExplSyntaxOn
\RenewDocumentCommand \emailauthor { m m }
  {\int_gincr:N \g_ead_int
    \seq_gput_right:Nn \g_stm_ead_seq
      {{ \ttfamily \tl_to_str:n { #1 }}
        \space ( #2 )}}
\RenewDocumentCommand \orcidauthor { m m }
  {\seq_gput_right:Nn \g_stm_orcid_seq
      {{ \ttfamily \tl_to_str:n { #1 }}
        \space ( #2 )}}
\ExplSyntaxOff
\makeatletter
\let\OriginalFirstPageNotes\printFirstPageNotes
\renewcommand{\printFirstPageNotes}{%
  \@dblfloat{figure}[!b]%
    \begin{minipage}{\textwidth}
      \def\thempfn{\thefootnote}%
      \OriginalFirstPageNotes
    \end{minipage}%
  \end@dblfloat
}
\makeatother

\begin{document}
\shorttitle{}
\shortauthors{Denaro and Lin}
\title[mode=title]{CODE-PIF: Convex/Deep (CODE) Multiview Pseudoinvariant Features for Multitemporal Multispectral Surface Reflectance Imagery}

\author[1]{Chia-Hsiang Lin}[orcid=0000-0002-4865-2329]
\ead{chiahsiang.steven.lin@gmail.com}
\fnmark[1]
\cormark[1]

\author[1,2]{Lino Garda Denaro}[orcid=0000-0001-5367-5392]
\ead{dlinogarda@brin.go.id}
\ead{dlinogarda@gmail.com}
\fnmark[1]

\author[1]{Jian-Kai Huang}[orcid=0009-0009-2748-7721]
\ead{q38141509@gs.ncku.edu.tw}

\affiliation[1]{organization={Department of Electrical Engineering, National Cheng Kung University (NCKU)},city={Tainan},postcode={70101},country={Taiwan}}
\affiliation[2]{organization={Research Center for Geoinformatics, National Research and Innovation Agency (BRIN)},country={Indonesia}}
\cortext[1]{Corresponding author.}
\fntext[equal]{The first two authors contributed equally to this work.}

\begin{abstract}
Multitemporal multispectral surface reflectance imagery exhibits radiometric and spectral variability due to atmospheric effects, sensor differences, and actual surface changes, complicating the identification of pseudoinvariant features (PIFs). 
The state-of-the-art deep multitemporal generalized canonical correlation analysis (DeepMGCCA) captures nonlinear dependencies across acquisition dates, but similarity in its learned feature space does not directly establish consistency in the original reflectance domain.
Meanwhile, we believe that the PIFs extracted by DeepMGCCA preserve essential structural characteristics, providing an effective prior to guide subsequent refinement.
Therefore, we propose the CODE-PIF architecture, an unsupervised convex/deep (CODE) framework that refines the coarse PIF from DeepMGCCA using convex optimization. 
Specifically, CODE-PIF uses the data-fitting term to integrate reflectance deviations, correlation-weighted spectral angles, and temporal-reference change information to promote radiometric and spectral consistency between predicted PIFs and observed data.
Moreover, the regularization term uses a mathematically simple $\bQ$-quadratic norm to incorporate information from the coarse PIF estimate into the refined solution.
The resulting convex optimization problem is solved using the alternating direction method of multipliers (ADMM), which decomposes it into simpler subproblems and provides convergence guarantees under standard assumptions.
Experiments across four regions of interest in Turkey, Taipei, Tainan, and Saudi Arabia show improved average radiometric and spectral agreement relative to DeepMGCCA. 
Averaged across these regions, the RMSE decreases from 0.0192 to 0.0164, PSNR increases from 35.91 to 37.44 dB, and the between-date spectral angle decreases from 3.77° to 2.97°. 
Although structural similarity and Pearson correlation do not improve in every scene, the results suggest that using observed reflectance data helps select PIFs with more consistent reflectance values and spectral shapes across dates.  

\end{abstract}

\begin{keywords}
Convex optimization (CO)\sep
Deep learning (DE)\sep
Alternating direction method of multipliers (ADMM) \sep
Deep multiview learning \sep
Pseudoinvariant features (PIFs) \sep
Multivariate alteration detection (MAD) \sep
Multitemporal surface reflectance harmonization \sep
Multispectral image
\end{keywords}
\maketitle
\section{Introduction}\label{sec:introduction}

In multitemporal optical remote sensing, achieving radiometric consistency is critical to ensure that observed reflectance differences primarily reflect actual surface changes rather than acquisition-dependent artifacts. 
Relative radiometric normalization commonly relies on pseudo-invariant features (PIFs), whose selection fundamentally governs the accuracy of subsequent calibration. 
Early evaluations of relative normalization methods established the importance of consistent reference observations \citep{Yuan1996,TIR2015,Fidelity1997}. 
Based on this foundation, invariant-pixel methods successfully connected radiometric normalization with cloud screening and spectrally consistent regression \citep{LIN2015107,SYARIZ201956}. 
Concurrently, studies focusing on bright and pseudo-invariant calibration targets highlighted the significance of material suitability \citep{BrightTargets2011,MIVIS2016}, while applications in ALOS normalization and variance-preserving forest analysis demonstrated the profound consequences of PIF selection for quantitative downstream tasks \citep{ALOS2012,Variance2015}. 
To address more complex relationships among multitemporal data, advanced techniques such as nonlinear cross-sensor spectral alignment using kernel canonical correlation analysis (CCA) \citep{Volpi2015}. 
In addition, the automatic MAD and IR-MAD algorithms \citep{CANTY2004441,CANTY20081025} have been introduced alongside foundational radiometric-control-set methods \citep{HALL199111,Nonlinear2016}. 
These methodological developments highlight the requirement to assess the spectral stability of candidate invariant pixels in the original reflectance domain, rather than relying solely on alignment in the transformed feature space.

Although consistency across image collections is usually addressed through global-to-local color balancing \citep{AGLCB2017}, guided optimization \citep{Guided2018}, and advanced adjustment techniques \citep{GlobalLocal2020,ProbColor2022,Block2024}, these methods primarily focus on visual consistency. 
Moreover, recent cross-sensor integration has demonstrated significant progress through spatiotemporal least-squares adjustments for PlanetScope \citep{LSSTAA2026}, Landsat-8/9 and Sentinel-2 interoperability \citep{Interop2024}, and joint harmonization and super-resolution frameworks \citep{Unmix2024,MultiImageSR2023,NormGAN2025,WEGLA2026}. 
However, true quantitative reflectance harmonization demands strict spectral preservation, rather than merely visually consistent or spatially enhanced results. 
To meet these rigorous quantitative requirements, invariant-feature methods have evolved considerably, as detailed in \citet{Hussain2013}. 
Specifically, multivariate alteration detection (MAD) leverages CCA-derived components to identify change \citep{NIELSEN19981}, with iterative reweighting further identifying likely invariant observations \citep{CANTY20081025}. 
By generalizing correlation analysis to multiple views, GCCA \citep{Horst_1961} has established a foundation for robust multitemporal PIF approaches \citep{Denarojpji2019,lin2019}. 
More recently, to accommodate highly complex cross-sensor relationships, researchers have transitioned toward nonlinear PIF regression \citep{Denaro2020} and kernel CCA \citep{Volpi2015}. 
Building on this, trainable Deep CCA/GCCA representations \citep{pmlrandrew13}, alongside reconstruction regularization \citep{pmlrwangb15}, feature mapping \citep{IterativeMapping2018}, and deep MAD frameworks \citep{Tu2024}, have emerged as scalable alternatives for comprehensive cross-source analysis.

Related research in multimodal change detection has explored graph-based modeling \citep{SimilarityGraphs2024,CGSL2025}, Fourier-domain structures \citep{Fourier2023}, domain adaptation \citep{DRUA2026}, and probabilistic modeling \citep{DeepProb2024}. 
In addition, deep learning for remote sensing encompasses content-invariant translation and patch exchange \citep{CIT2023,I3PE2023}, convolutional modeling and supervised fusion \citep{CLNet2021,DFN2020}, and attention mechanisms \citep{IWDQUEEN,tang2025quantum}. 
On the other hand, quantum deep learning architectures have also been investigated for hyperspectral remote sensing tasks, providing complementary features derived from quantum unitary transformations that can enhance the performance of conventional neural networks \citep{HyperKING,QDIP,HyperQUEEN}.
Recently, the connection between calibration and change detection has become more explicit, which uses unsupervised, incremental, safe pseudo-labeling to identify unchanged samples and progressively estimate a spectral calibration function \citep{HyperLUCID}.
Although these methods provide diverse feature representations, their effectiveness in other remote sensing tasks does not necessarily imply reliable PIF selection for radiometric calibration. 

To address these challenges, we propose CODE-PIF, an unsupervised convex/deep (CODE) framework \citep{CODE} for improving pseudoinvariant feature (PIF) selection in multitemporal surface reflectance imagery.
The CODE learning criterion was initially developed for hyperspectral remote sensing and subsequently extended to semiconductor and artificial intelligence applications \citep{10047994,NCCODE}. 
By integrating convex optimization (CO) with deep learning (DE), CODE incorporates a rough deep prediction as prior information within a regularized convex optimization problem, combining the observed data and the initial deep rough prediction to make the final PIFs.
Specifically, the data-fitting term ensures consistency with observed data, enabling refinement of the initial PIFs without an additional correction network. 
Moreover, the mathematically simple $\bQ$-quadratic-norm regularization term serves as a feature extractor to extract the embedded features in rough PIFs, and then fuse them into the final target PIFs.
Therefore, CODE-PIF first uses DeepMGCCA \citep{DenaroLin2026Monotone} to extract an initial set of PIFs that retain valuable structural information to guide subsequent improvements.
The architecture and training procedure of DeepMGCCA are detailed in that work, whereas the present study focuses on refining the resulting PIF selection through convex optimization.
This refinement is motivated by the fact that agreement in the learned latent space does not necessarily imply radiometric and spectral consistency in the original reflectance observations.
Moreover, converting continuous feature scores into a binary mask inherently discards the relative ranking of candidates. 
Since incorrectly selected reference pixels can bias the estimated radiometric transformation, CODE-PIF incorporates coarse prior information from the observed reflectance data to refine PIF selection.

Based on the CODE frameworks \citep{CODEIF,QRCODE}, CODE-PIF constructs a robust invariance score directly from the observed reflectance data for each acquisition date. 
The underlying change identifications combine correlation-weighted spectral angles and deviations from a temporal reference. 
While spectral angles characterize shifts in spectral direction, the temporal-reference term quantifies absolute discrepancies in reflectance values. 
In addition, these complementary measures capture physical evidence that extends well beyond the latent-space correlations extracted by DeepMGCCA \citep{DenaroLin2026Monotone}. 
This spectral-angle-based fidelity \citep{CODEHCD} is then fused with the initial deep invariant mask using a well-designed convex optimization problem. 
Then, we employ the mathematically simple $\bQ$-quadratic-norm regularization term \citep{CODE} to fuse the rough solution from DeepMGCCA into the final PIFs. 
Compared with total variation \citep{TV} or an explicit self-similarity \citep{S4} function, the $\bQ$-quadratic-norm regularizer offers a mathematically simple and differentiable formulation.
Consequently, CODE-PIF can reduce the scores of initially selected pixels whose spectral angles differ substantially from the reference and increase the scores of spectrally consistent pixels that DeepMGCCA did not initially select, thereby improving the final PIF selection.

For each acquisition date, we solve the corresponding convex optimization problem using the alternating direction method of multipliers (ADMM) \citep{ChiConvex2017}, which has been applied to a variety of remote sensing tasks \citep{PAINT2026}.
ADMM combines the decomposability of dual ascent with the convergence properties of the method of multipliers.
This structure allows the optimization problem to be decomposed into simpler subproblems while retaining convergence guarantees under standard assumptions.
The optimized continuous invariance scores subsequently guide the selection of PIFs for radiometric harmonization.
The reflectance quantiles of the selected PIFs \citep{DenaroLin2026Monotone} are used to estimate a monotone piecewise-linear transformation for each acquisition date. 
CODE-PIF selects the calibration references, while the transformation adjusts reflectance values without reversing their ordering.
This procedure aims to reduce acquisition-dependent differences among stable targets while preserving genuine surface changes, thereby supporting more consistent multitemporal comparisons.
The main contributions of this work are summarized as follows. 
\begin{enumerate}
\item We propose a convex/deep framework \citep{CODE}, CODE-PIF, for refining DeepMGCCA-derived PIF selection \citep{DenaroLin2026Monotone} by integrating the initial deep mask with invariance scores derived from correlation-weighted spectral angles and temporal-reference deviations.
The resulting continuous scores enable both re-ranking of initially selected pixels and consideration of candidates excluded by the initial PIFs from DeepMGCCA.
\item We evaluate CODE-PIF on Landsat-8/9 and Sentinel-2 time series data spanning urban, desert, coastal, and forest environments, using the refined PIFs as reference targets for monotone piecewise-linear radiometric harmonization.
The proposed CODE-PIF achieves state-of-the-art performance across these four geographically distinct regions, supporting its applicability to diverse environmental settings.
\item The convex formulation of CODE-PIF provides a mathematical foundation for developing an interpretable deep unfolding architecture \cite{COSUP} for multitemporal surface reflectance harmonization, where model interpretability is important for downstream applications \citep{HyperLUCID}. 
Future work will focus on unfolding the associated iterative optimization algorithm into trainable network stages, combining the explicit structure of the optimization model with the flexibility of deep learning.
\end{enumerate}

The remainder of this paper is organized as follows.
Section~\ref{sec:meth} presents the mathematical formulation, optimization algorithm, and radiometric harmonization procedure of CODE-PIF.
Section~\ref{sec:experiment} demonstrates comprehensive qualitative and quantitative evaluations of the proposed framework.
Section~\ref{sec:conclusion} concludes the article and summarizes the future work of this study.

\section{Methodology}\label{sec:meth}

\begin{figure*}[t]
\centering
\draftincludegraphics[width=0.95\linewidth]{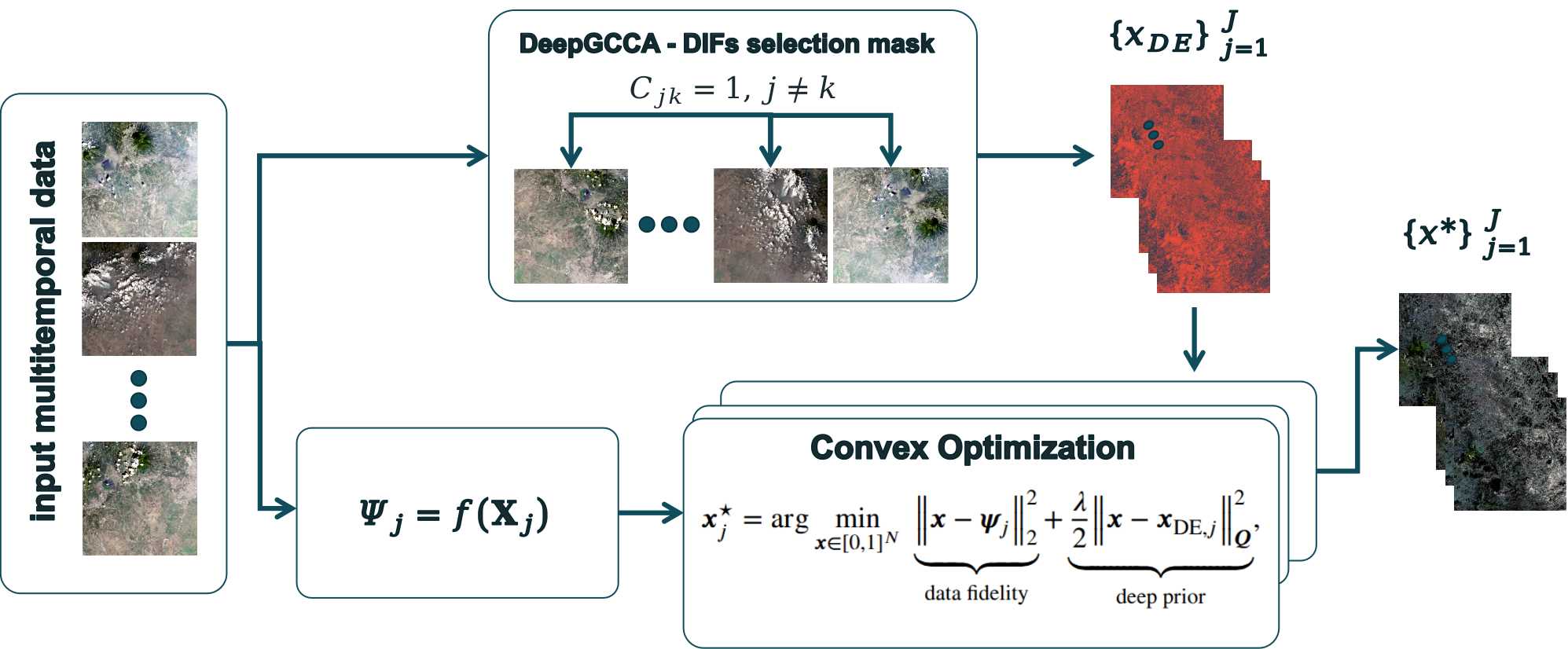}{0.95\linewidth}
\caption{Overview of the proposed CODE-PIF framework. 
Given co-registered multitemporal surface reflectance images $\mathbf{X}_j$, $j=1,\ldots,J$, DeepMGCCA uses all date pairs ($C_{jk}=1$ for $j\neq k$) to identify the initial PIF mask $\bm{\hat{x}}_{DE_j}$. 
This mask defines the deep change prior $\bx_{DE,j}=\mathbf{1}-\bm{\hat{x}}_{\text{DE}_j}$, which assigns zero to initially selected PIFs and one elsewhere. 
A complementary change map $\bpsi_j$ is computed from the multitemporal reflectance observations. 
For each date, a strictly convex objective combines this map with the deep prior through a data-fitting term and a $\bQ$-quadratic-norm regularizer, obtaining the continuous change scores $\bx_j^\star$. The lowest-scoring fraction $\tau$ is selected as the final PIF population for subsequent harmonization. 
The stacked maps represent separately optimized outputs for individual acquisition dates.}
\label{fig:general}
\end{figure*}

Let $\mathbf{X}_j\in\mathbb{R}^{B\times N}$,
$j=1,\ldots,J$, denote the co-registered
multispectral surface reflectance imagery
acquired at date $j$, where $B$ and $N$
represent the numbers of spectral bands
and pixels, respectively. 
Let $\mathcal{V}\subseteq\{1,\ldots,N\}$
denote the set of pixels valid across all
acquisition dates. 
As shown in Fig.~\ref{fig:general}, the CODE-PIF framework comprises two main steps. 
First, DeepMGCCA \citep{DenaroLin2026Monotone} learns multiview features and identifies the initial PIFs $\bm{\hat{x}}_{\text{DE}_j}\in\{0,1\}^N$ for each date $j$ (cf. Section~\ref{sec:deep}). 
Subsequently, the CODE formulation integrates the DeepMGCCA-derived prior with spectral and temporal information from the observed reflectance data through convex optimization, producing the refined invariance scores (cf. Section~\ref{sec:code}).

\subsection{DeepMGCCA Deep Prior}\label{sec:deep}

The initial PIF mask is obtained using DeepMGCCA \citep{DenaroLin2026Monotone}, which extends linear multiview invariant-feature learning to a nonlinear setting to model complex spectral-spatial relationships and radiometric variations without requiring ground-truth PIF labels.
DeepMGCCA employs a shared encoder to capture spectral-spatial structure common across acquisition dates, while date-specific heads account for variations caused by acquisition.
After learning these representations, the DeepMGCCA applies multivariate alteration analysis to pairs of connected data. 
Pixels with consistent canonical representations across dates receive low alteration scores, whereas differences associated with land-cover changes, illumination variations, or other radiometric differences can lead to higher scores. 
Consequently, the resulting scores provide a deep, data-driven prior for identifying pseudo-invariant pixels without assuming that radiometric differences are purely linear, satisfying real-world scenarios. 

In this study, pixels are ranked by their normalized alteration scores for each acquisition date $j$, and the lowest-scoring fraction $\tau$ is selected to construct the initial PIF mask $\hat{\bm{x}}_{\text{DE}_j}$. 
CODE-PIF uses this binary mask, rather than the learned latent representations, to define the deep change prior $\bm{x}_{\text{DE}_j}=\mathbf{1}-\hat{\bm{x}}_{\text{DE}_j}$.
The resulting prior assigns $0$ to pixels selected as PIFs by DeepMGCCA and $1$ to all remaining pixels, providing an initial estimate for subsequent refinement through convex optimization.
The complete architecture, optimization procedure, and statistical derivation of DeepMGCCA are provided by \citet{DenaroLin2026Monotone}.

\subsection{Convex Framework}\label{sec:code}
Based on the initial PIF estimate provided by DeepMGCCA (cf. Section~\ref{sec:deep}), we introduce spectral and temporal consistency measures derived from the observed reflectance data.
We then formulate a convex optimization problem that integrates these measures with the deep prior through a data-fitting term and a $\bQ$-quadratic-norm regularization term \citep{CODE}.
Subsequently, we derive the corresponding ADMM update rules and explain how the optimized scores guide the final PIF selection.
Finally, the selected PIFs serve as reference targets for estimating monotone piecewise-linear transformations to harmonize the multitemporal reflectance observations. 
First, we can mathematically define the criterion for each date $j$ as
\begin{equation}
\bx_j^\star
=
\arg\min_{\bx\in[0,1]^N}
\operatorname{DF}_j(\bx)
+
\lambda\operatorname{REG}_j(\bx),
\quad
\forall j=1\dots J,
\label{eq:code_general}
\end{equation}
where $\lambda>0$ controls the regularization strength.
The vector $\bx$ contains continuous change scores, with smaller values indicating stronger support for PIF selection, while the data-fitting term $\operatorname{DF}_j$ relates these scores to spectral and temporal information derived from the observations.
The regularizer $\operatorname{REG}_j$ incorporates the initial selection obtained from DeepMGCCA \citep{DenaroLin2026Monotone}.

Specifically, the data-fitting term promotes consistency between the refined PIF scores and information derived from the observed reflectance data.
The consistency is important for reliable downstream analysis, as favorable evaluation metrics alone do not guarantee radiometric and spectral fidelity. 
Therefore, we incorporate the observed temporal information and spectral angle measurements into the data-fitting term to promote radiometric and spectral consistency in the final PIF selection, which is detailed in Section \ref{sec:data_fitting}
Moreover, to incorporate useful information from the coarse PIF selection from DeepMGCCA into the refined solution, we adopt a mathematically simple $\bQ$-quadratic-norm regularizer \citep{CODE}, which can be defined as
\begin{equation}
\operatorname{REG}_j(\bx)
\triangleq
\frac{1}{2}
\big\lVert
\bx-\bx_{\mathrm{DE},j}
\big\rVert_{\bQ}^{2},
\label{eq:regularizer}
\end{equation}
where $\bx_{\mathrm{DE},j}$ is the deep change prior derived from the DeepMGCCA \citep{DenaroLin2026Monotone}. 
For any vector $\bm{v}\in\mathbb{R}^{N}$ and positive semidefinite matrix $\bQ\in\mathbb{R}^{N\times N}$, we define
$\lVert\bm{v}\rVert_{\bQ}^{2}\triangleq\bm{v}^{T}\bQ\bm{v}$.
Compared with the nonsmooth total variation (TV) regularizer and graph-based self-similarity regularizer \citep{TV,S4}, the $\bQ$-quadratic regularizer provides a simple, differentiable formulation to facilitate efficient optimization.
The data-fitting term is presented in Section~\ref{sec:data_fitting}, and the resulting criterion is presented and solved in Section~\ref{sec:admm}.

\subsection{Multitemporal Spectral Data-Fidelity Map}\label{sec:data_fitting}
In the following part, we construct a multitemporal spectral data-fidelity map from the observed reflectance data.
The design combines the differences in spectral angle across acquisition dates with reflectance residuals weighted by a temporal reference.
Specifically, spectral angles characterize differences in spectral shape, whereas the temporal reference integrates repeated observations to guide the construction of a scene baseline and the weighting of bandwise residuals.
Therefore, we can define the temporal reference as
\begin{equation}
T_{b,\ell}
=
\min\!\left[
\frac{1}{J}\sum_{j=1}^{J}\mathbf{X}_{j,b,\ell},
\operatorname{median}_{1\leq j\leq J}\left(\mathbf{X}_{j,b,\ell}\right)
\right],
\label{eq:tref}
\end{equation}
where $b$ and $\ell$ represent $b$-th band and $\ell$-th pixel, respectively. 
This construction is designed to reduce sensitivity to isolated bright observations, such as those affected by clouds.
These observations can significantly increase the temporal mean, whereas the median is generally less sensitive to a small number of high-valued outliers. 
Therefore, selecting the smaller of these statistics limits the influence of these observations.

Let $\theta_{jk,\ell}\in \mathbb{R}$ denotes the spectral angle between the observed reflectance vectors $\mathbf{X}_{j,\ell}$ and $\mathbf{X}_{k,\ell}$.
To account for differences in cross-date agreement, let $\rho_{jk}\in[0,1]$ denote a weight derived from the band-averaged Pearson correlation between dates $j$ and $k$.
The weighted spectral-angle score and its normalized counterpart are defined as
\begin{equation}
\Theta_{j,\ell}
=
\frac{\sum_{k\neq j}\rho_{jk}\theta_{jk,\ell}}
{\sum_{k\neq j}\rho_{jk}},
\quad
\tilde{\Theta}_{j,\ell}
=
\frac{\Theta_{j,\ell}}
{\displaystyle
\max_{\substack{j^{\prime},\ell^{\prime}\in\mathcal{V}}}
\Theta_{j',\ell'}}.
\label{eq:sam}
\end{equation}
The correlation weights emphasize date pairs with higher reflectance correlation over pixels initially selected as PIFs in both images.
Normalization uses a common maximum across all dates and pixels to preserve the relative scale of spectral deviations and avoid amplifying small differences.

\begin{algorithm}[t]
\caption{CODE-PIF Algorithm}
\label{alg:code}
\begin{algorithmic}[1]

\STATE \textbf{Given} multitemporal MSIs
$\mathbf{X}_1,\ldots,\mathbf{X}_J$, valid-pixel set $\mathcal{V}$,
selection fraction $\tau$, parameters $\lambda,c>0$,
tolerance $\epsilon$, and maximum iterations $T_{\max}$.

\STATE Obtain the initial PIF masks
$\{\bm{\hat{x}}_{\text{DE}_j}\}_{j=1}^{J}$
using DeepMGCCA with selection fraction $\tau$.

\STATE Compute the multitemporal data-fidelity maps
$\{\bpsi_j\}_{j=1}^{J}$
using \eqref{eq:tref}--\eqref{eq:psi}.

\FOR{$j=1,\ldots,J$}

    \STATE Compute the deep change prior
    $\bx_{DE,j}=\mathbf{1}-\bm{\hat{x}}_{\text{DE}_j}$.

    \STATE Initialize
    $\bx^0=\bz^0=\bu^0=\mathbf{0}$ and $t=0$.

    \REPEAT
        \STATE Update $\bz^{t+1}$ using \eqref{eq:zstep}.
        \STATE Update $\bx^{t+1}$ using \eqref{eq:xstep}.
        \STATE Update $\bu^{t+1}$ using \eqref{eq:ustep}.
        \STATE $t:=t+1$.
    \UNTIL the predefined stopping criterion is met

    \STATE Set $\bs_j=\bx^\star$ and
    $s_{j,\ell}=1$ for $\ell\notin\mathcal{V}$.

    \STATE Select the $n_j$ pixels with the lowest scores from
    $\{\ell\in\mathcal{V}:s_{j,\ell}<1\}$,
    breaking ties using a fixed random seed.

    \STATE Construct the final PIF mask
    $\bm{\hat{x}}_{\mathrm{CODE},j}$,
    assigning $1$ to selected pixels and $0$ elsewhere.

\ENDFOR

\STATE \textbf{Output} $\{\bm{\hat{x}}_{\mathrm{CODE},j}\}_{j=1}^{J}$.

\end{algorithmic}
\end{algorithm}

To emphasize spectrally consistent pixels when estimating the weighted mean reflectance, we define the invariance weight as $\omega_{j,\ell}=1-\tilde{\Theta}_{j,\ell}$.
The weighted scene mean and the reference-weighted residual are then given by
\begin{equation}
m_{j,b}
=
\frac{
\sum_{\ell\in\mathcal{V}}
\omega_{j,\ell}T_{b,\ell}\mathbf{X}_{j,b,\ell}
}{
\sum_{\ell\in\mathcal{V}}
\omega_{j,\ell}T_{b,\ell}
},
\mkern1mu
\tilde{X}_{j,b,\ell}
=
\frac{T_{b,\ell}\left(\mathbf{X}_{j,b,\ell}-m_{j,b}\right)}
{\sum_{b'=1}^{B}T_{b',\ell}},
\label{eq:center}
\end{equation}
where $m_{j,b} \in \mathbb{R}$ is the weighted mean reflectance over valid pixels, and $\tilde{X}_{j,b,\ell}$ measures deviation from this mean.
The temporal reference weights both the mean estimation and the bandwise residuals.

Finally, the resulting data-fidelity map combines the normalized spectral angle score with the residual energy,
\begin{equation}
\hat{\psi}_{j,\ell}
=
\tilde{\Theta}_{j,\ell}
\sum_{b=1}^{B}\tilde{X}_{j,b,\ell}^{2},
\quad
\psi_{j,\ell}
=
\frac{\hat{\psi}_{j,\ell}}
{\displaystyle
\max_{\substack{j^{\prime},\ell^{\prime}\in\mathcal{V}}}
\hat{\psi}_{j',\ell'}},
\label{eq:psi}
\end{equation}
where $\psi_{j,\ell}\in[0,1]$, and invalid pixels are assigned $1$.
The score combines spectral shape differences with weighted deviations from the mean reflectance.
Identical spectral shapes give zero cost before normalization, even if reflectance values differ.
Thus, $\bm{\psi}$ is a relative change score for convex optimization, not a probability of actual surface change.

\subsection{Algorithm Derivation using ADMM}\label{sec:admm}
Using the multitemporal spectral angle data-fidelity map $\bpsi_j$ in \eqref{eq:psi}, we define the data-fitting term as $\operatorname{DF}_j(\bx)\triangleq\lVert\bx-\bpsi_j\rVert_2^2$. 
By incorporating the data-fitting term into \eqref{eq:code_general}, we can derive the CODE formulation for each date $j$ as 
\begin{equation}
\bx_j^\star
=
\arg\min_{\bx\in[0,1]^N}
\big\lVert\bx-\bpsi_j\big\rVert_2^2
+
\frac{\lambda}{2}
\big\lVert
\bx-\bx_{\mathrm{DE},j}
\big\rVert_{\bQ}^2,
\label{eq:code}
\end{equation}
where $\bx_j^\star\in[0,1]^N$ is the final PIF scores.
Before solving \eqref{eq:code}, we specify the positive semidefinite matrix $\bQ$ in the regularization term.
Following previous articles \citep{CODEHCD,CODEMM}, $\bQ$ can be constructed as $\bQ=\bE\bE^T$, where the columns of $\bE$ form an orthonormal basis for a feature subspace using PCA \citep{CODE}. 
This formulation explicitly penalizes the components of the deviation from the deep rough PIFs that reside within that subspace.
For the single band change score map $\bx_j^\star$, the matrix $\bQ$ can be simplified to the identity matrix $\bI_N$ in this work.
Therefore, \eqref{eq:code} can be reformulated as
%
\begin{equation}
\min_{\bx}
\Big[
\underbrace{
\bx^T(\bI-\bm{D_j})\bx
}_{\text{invariant push}}
+
\underbrace{
(\mathbf1-\bx)^T\mathbf \bm{D_j}(\mathbf1-\bx)
}_{\text{changed push}}
+
\frac{\lambda}{2}
\big\lVert\bx-\bx_{\text{DE},j}\big\rVert_{2}^{2}
\Big],
\label{eq:code_push}
\end{equation}
where $\bm{D_j}=\operatorname{Diag}(\bpsi_j)$, and $\mathbf{1}\in\mathbb{R}^N$ is the all-ones vector. 
The two complementary observation terms are designed to support adjustments to the initial PIF selection in both directions.
Pixels with low observed change scores are favored as invariant candidates, whereas those with high scores are discouraged, allowing both previously excluded candidates and potentially inconsistent initial selections to be reconsidered.
The regularizer preserves guidance from the DeepMGCCA prior \citep{DenaroLin2026Monotone} while allowing the observed reflectance consistency to influence the final selection. 

To solve \eqref{eq:code_push} using ADMM solver \citep{CODE}, we introduce an auxiliary variable $\bz\in\mathbb{R}^N$, and \eqref{eq:code_push} can be reformulated as 
\begin{equation}
\begin{aligned}
\min_{\bx=\bz}\;&
\big\lVert\bx-\bpsi_j\big\rVert_2^2
+
\frac{\lambda}{2}
\big\lVert
\bz-\bx_{\mathrm{DE},j}
\big\rVert_{2}^2, 
\end{aligned}
\label{eq:admm_split}
\end{equation}
where each entry of $\bx$ is constrained to lie within the interval $[0,1]$.
This splitting assigns the deep-prior penalty to $\bz$ and retains the data-fitting term and score bounds on $\bx$.
Subsequently, the augmented Lagrangian of \eqref{eq:admm_split} can be derived as
\begin{equation}
\mathcal{L}(\bx,\bz,\bu)
={}
\big\lVert\bx-\bpsi_j\big\rVert_2^2
+
\frac{\lambda}{2}
\big\lVert
\bz-\bx_{\mathrm{DE},j}
\big\rVert_{2}^2
+
\frac{c}{2}
\big\lVert
\bx-\bz+\bu
\big\rVert_2^2,
\label{eq:admm_lagrangian}
\end{equation}
where $c>0$ represents the penalty parameter and $\bu\in\mathbb{R}^N$ denotes the scaled dual variable. 

By removing terms independent of $\bz$, we can update the variable $\bz$ as follows.
\begin{equation}
\begin{aligned}
\bz^{t+1}
&=\arg\min_{\bz}
\frac{\lambda}{2}
\big\lVert\bz-\bx_{\mathrm{DE},j}\big\rVert_2^2+
\frac{c}{2}
\big\lVert\bz-(\bx^t+\bu^t)\big\rVert_2^2
\\
&=\frac{
\lambda\bx_{\mathrm{DE},j}
+c(\bx^t+\bu^t)
}{\lambda+c}.
\end{aligned}
\label{eq:zstep}
\end{equation}
Similarly, the updating rule of variable $\bx$ is as
\begin{equation}
\begin{aligned}
\bx^{t+1}
&=\arg\min_{\bx\in[0,1]^N}
\big\lVert\bx-\bpsi_j\big\rVert_2^2+
\frac{c}{2}
\big\lVert\bx-(\bz^{t+1}-\bu^t)\big\rVert_2^2
\\
&=\Pi_{[0,1]^N}\!\left(
\frac{
2\bpsi_j+c(\bz^{t+1}-\bu^t)
}{2+c}
\right),
\end{aligned}
\label{eq:xstep}
\end{equation}
where $\Pi_{[0,1]^N}$ denotes the componentwise projection onto $[0,1]$, defined by
$[\Pi_{[0,1]^N}(\bm{v})]_\ell
=\min\{1,\max\{0,v_\ell\}\}$.
Finally, the scaled dual variable is updated as
\begin{equation}
\bu^{t+1}
=
\bu^t+\bx^{t+1}-\bz^{t+1}.
\label{eq:ustep}
\end{equation}
We initialize $\bx^0$, $\bz^0$, and $\bu^0$ to zero, and these updates are repeated until the maximum iteration $T_{\max}$ is reached.
The final iterate provides the refined change scores for subsequent PIF selection, with invalid pixels assigned a score of $1$. 
The detailed updating rule of CODE-PIF is given in Algorithm~\ref{alg:code}.

\begin{table*}[htbp]
\centering
\caption{Overview of the four regions of interest (ROIs) used to evaluate CODE-PIF, including acquisition dates, sensors, processing resolutions, and environmental characteristics.}
\label{tab:roi_data}
\renewcommand{\arraystretch}{1.1}
\begin{tabular}{c c l l l c l}
\hline
\textbf{No.} & \textbf{ROI} & \textbf{Location}
& \textbf{Date} & \textbf{Sensor}
& \textbf{Resolution} & \textbf{Description} \\
\hline

\multirow{4}{*}{1} &
\multirow{4}{*}{
\draftincludegraphics[width=1.3cm]
{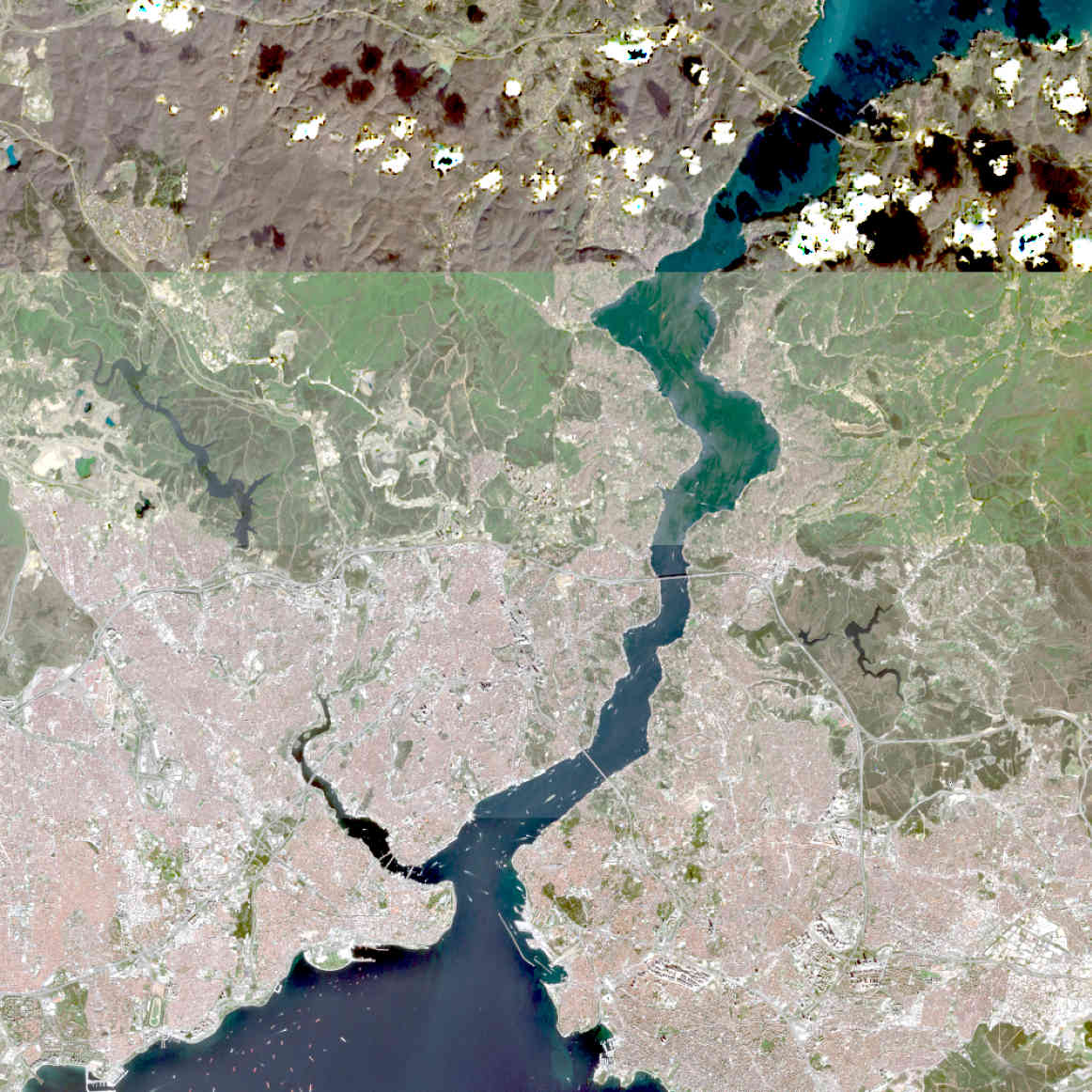}{1.3cm}} &
\multirow{4}{*}{Turkey}
& 2024-06-27 & Landsat 8
& \multirow{4}{*}{30 m}
& \multirow{4}{*}{City and Strait} \\
& & & 2024-04-16 & Landsat 9 & & \\
& & & 2024-03-24 & Sentinel-2A & & \\
& & & 2024-10-05 & Sentinel-2B & & \\
\hline

\multirow{4}{*}{2} &
\multirow{4}{*}{
\draftincludegraphics[width=1.3cm]
{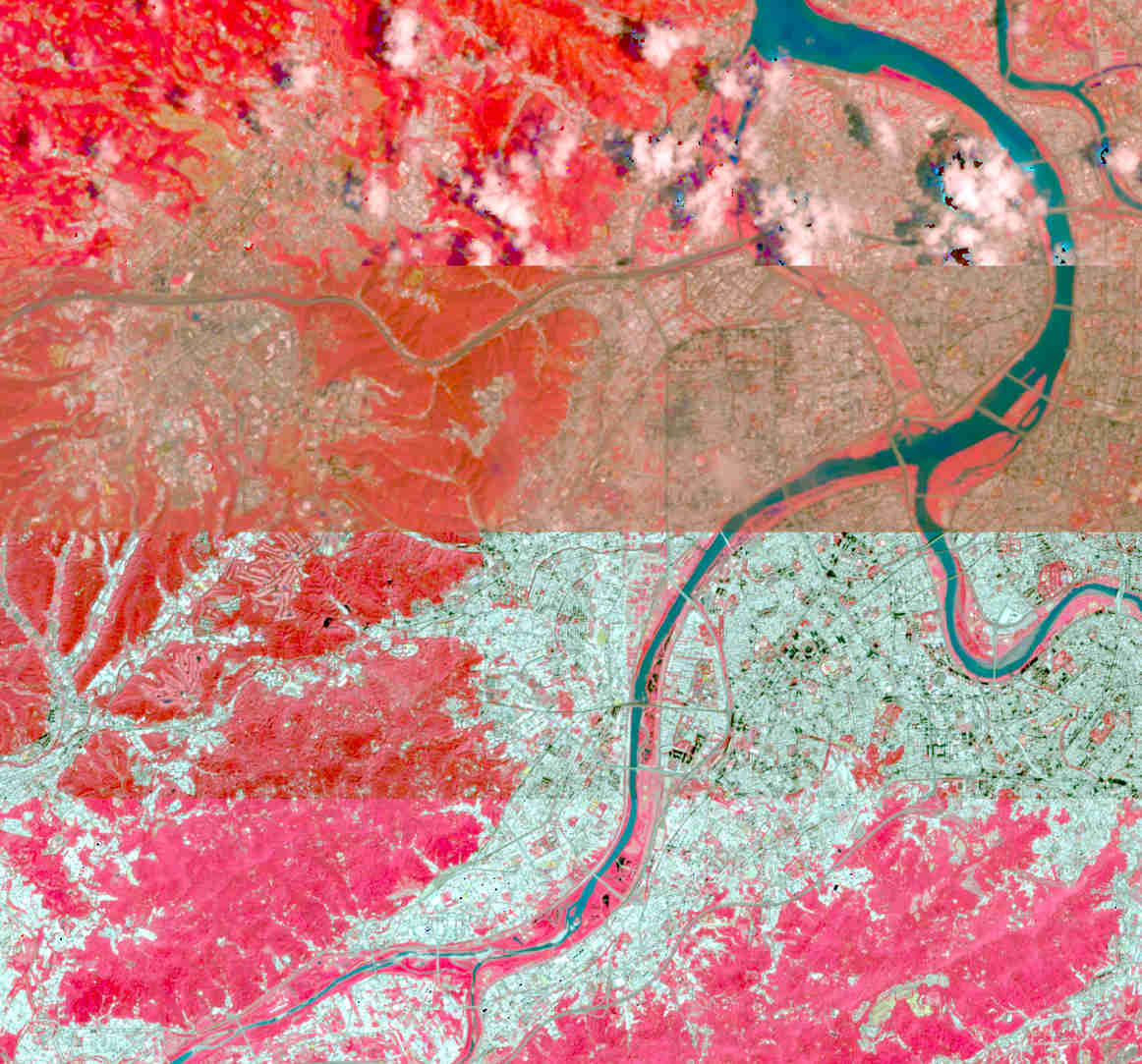}{1.3cm}} &
\multirow{4}{*}{Taipei}
& 2024-08-04 & Landsat 8
& \multirow{4}{*}{10 m}
& \multirow{4}{*}{Urban and River} \\
& & & 2024-03-21 & Landsat 9 & & \\
& & & 2024-02-14 & Sentinel-2B & & \\
& & & 2024-09-01 & Sentinel-2B & & \\
\hline

\multirow{4}{*}{3} &
\multirow{4}{*}{
\draftincludegraphics[width=1.3cm]
{fig/exp/T1tainan.jpg}{1.3cm}} &
\multirow{4}{*}{Tainan}
& 2024-04-21 & Landsat 8
& \multirow{4}{*}{30 m}
& \multirow{4}{*}{Urban and Mountainous} \\
& & & 2024-07-02 & Landsat 9 & & \\
& & & 2024-04-04 & Sentinel-2B & & \\
& & & 2024-12-20 & Sentinel-2B & & \\
\hline

\multirow{4}{*}{4} &
\multirow{4}{*}{
\draftincludegraphics[width=1.3cm]
{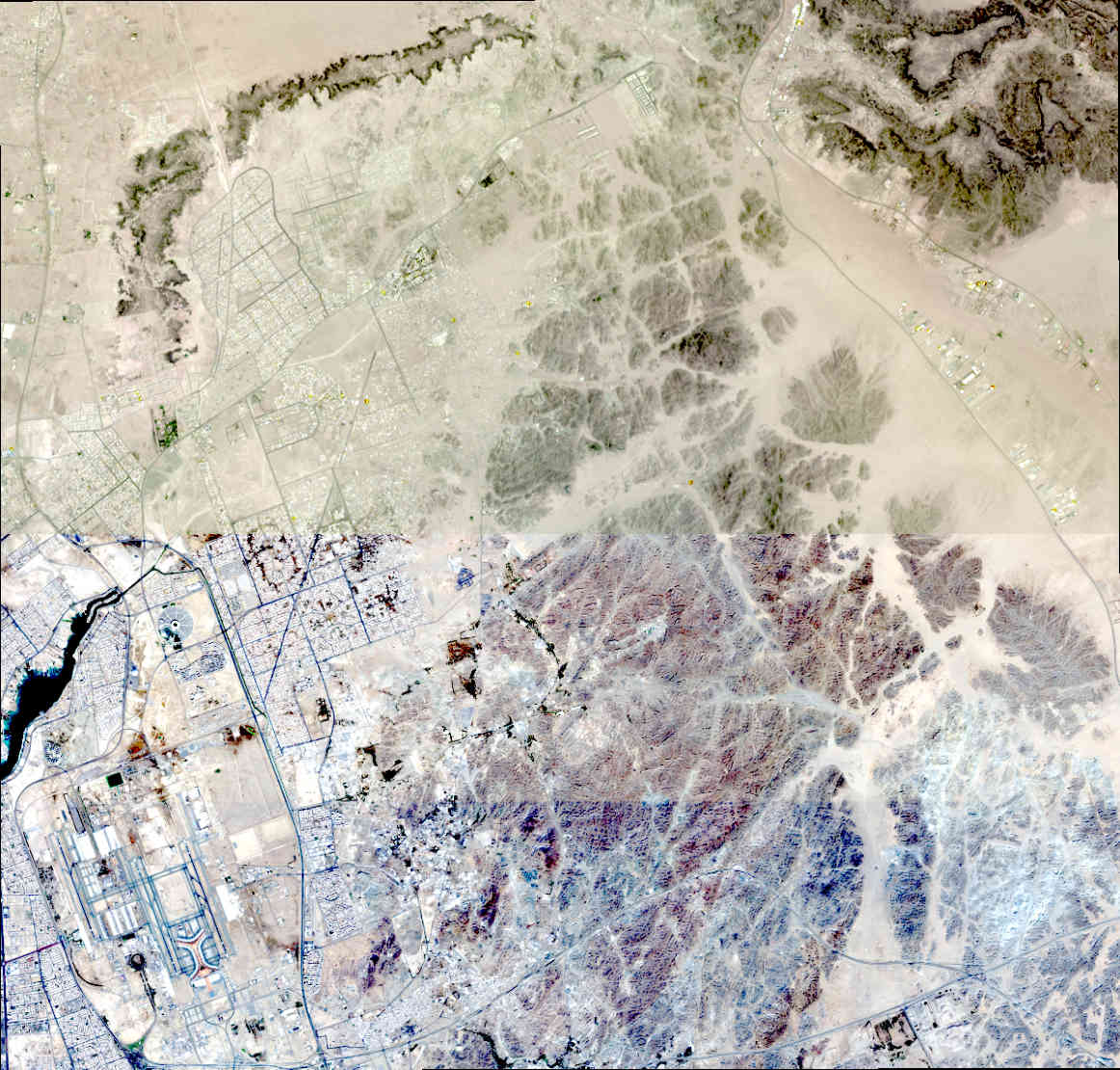}{1.3cm}} &
\multirow{4}{*}{Saudi Arabia}
& 2024-06-20 & Landsat 8
& \multirow{4}{*}{30 m}
& \multirow{4}{*}{City and Desert} \\
& & & 2024-04-25 & Landsat 9 & & \\
& & & 2024-11-06 & Sentinel-2B & & \\
& & & 2024-12-06 & Sentinel-2B & & \\
\hline
\end{tabular}
\end{table*}

\section{Experimental Results and Discussion}\label{sec:experiment}
This section evaluates the effectiveness of CODE-PIF in selecting radiometrically and spectrally consistent PIFs from multitemporal multispectral surface reflectance imagery. 
We compare CODE-PIF with DeepMGCCA using the same PIF selection fraction to assess the contribution of convex refinement. 
Quantitative evaluation uses RMSE, PSNR, SSIM, Pearson's correlation coefficient, and SAM, complemented by qualitative comparisons. 
These metrics are interpreted jointly to characterize radiometric agreement, structural similarity, and spectral stability, and to identify potential trade-offs across scenes.

\subsection{Datasets}
\label{sec:data}

Four geographically distinct ROIs, namely Turkey, Taipei, Tainan, and Saudi Arabia, were selected to evaluate the robustness of CODE-PIF under diverse geographic and land-cover conditions. 
These datasets provide contrasting spatial and spectral characteristics to assess multiview PIFs selection in multitemporal, multispectral surface reflectance imagery. 
Table~\ref{tab:roi_data} summarizes the geographic characteristics, acquisition dates, sensors, and processing resolutions of the four ROIs.
In Turkey, the study area comprises urban and coastal landscapes, including built-up surfaces, water bodies, and transitions between land and the strait.
The Taipei ROI encompasses heterogeneous urban and riverine areas, where built-up surfaces and water bodies exhibit spatially varying reflectance characteristics. 
The Tainan dataset captures urban, coastal, and mountainous environments with diverse surface materials and topographic variations.
For Saudi Arabia, the selected area represents an arid landscape characterized by extensive bare surfaces and urban development.
Furthermore, these environments enable evaluation of radiometrically and spectrally stable pixel selection across diverse landscape conditions.
Specifically, each ROI comprises four multispectral observations acquired in 2024 by Landsat-8/9 and Sentinel-2. 
For Turkey, the Landsat-8/9 images were acquired on June 27 and April 16, respectively, while the Sentinel-2A and Sentinel-2B images were acquired on March 24 and October 5, respectively. 
For the Taipei scenario, Landsat-8/9 acquisitions were obtained on August 4 and March 21, respectively, along with two Sentinel-2B observations on February 14 and September 1. 
In the case of Tainan, the Landsat 8 and Landsat 9 images were acquired on April 21 and July 2, respectively, accompanied by Sentinel-2B observations from April 4 and December 20. 
Moreover, the Landsat-8/9 acquisitions of Saudi Arabia were obtained on June 20 and April 25, respectively, while the two Sentinel-2B observations were acquired on November 6 and December 6.

The Turkey, Tainan, and Saudi Arabia datasets were processed on a common 30 m spatial grid, whereas the Taipei dataset used a 10 m grid comprising pansharpened Landsat imagery and spatially aligned Sentinel-2 observations. 
These processing resolutions specify the pixel spacing of the multitemporal analysis grids and should be distinguished from the native spatial resolutions of individual sensor bands. 
Within each ROI, all acquisitions were spatially co-registered and organized into a common multiview representation to enable pixelwise comparisons across dates.
Moreover, reliable multitemporal PIF selection requires both spatial alignment and a consistent valid-pixel domain. 
The Cloud and shadow screening strategies have been reviewed by \citet{CloudReview2022} and \citet{CloudShadow2018}, while \citet{Registration2016} investigated the geometric alignment of multitemporal optical imagery using stable line features. 
In this study, CODE-PIF takes the co-registered multispectral surface reflectance arrays and their associated validity masks as inputs. 
Therefore, PIF selection is restricted to pixels that satisfy the common validity requirements across all acquisition dates.

For each acquisition date, six approximately corresponding spectral bands are used, including blue, green, red, NIR, SWIR-1, and SWIR-2. 
These bands are drawn from Landsat-8/9 Operational Land Imager (OLI) and Sentinel-2 multispectral instrument observations, with their spectral characteristics summarized in Table~\ref{tab:bands}. 
Although the selected bands cover comparable spectral regions, differences in sensor spectral response functions may introduce cross-sensor reflectance discrepancies. 
Therefore, these differences should be considered when interpreting spectral consistency across acquisition dates.
Across the four datasets, the PIF populations selected by DeepMGCCA are compared with those refined by CODE-PIF using the same prescribed selection fraction. 
This setting supports a controlled evaluation of PIF selection quality using radiometric and spectral consistency criteria computed from corresponding multitemporal reflectance observations at the selected pixel locations. 
The evaluation thus focuses on the characteristics of the identified PIF populations rather than the performance of subsequent image transformations.

\begin{table}[t]
\caption{Detailed information on the Landsat 8/9 OLI and Sentinel-2 MSI spectral bands used in this study. 
All wavelengths are expressed in nanometers (nm).}
\label{tab:bands}
\renewcommand{\arraystretch}{1.2}
\begin{center}
\footnotesize
\setlength{\tabcolsep}{1.0mm}
\begin{tabular}{l c c c c c c}
\hline\hline
 & \multicolumn{3}{c}{\textbf{Landsat 8/9 OLI}} & \multicolumn{3}{c}{\textbf{Sentinel-2 MSI}} \\
\cmidrule(lr){2-4}\cmidrule(lr){5-7}
\textbf{Band} & \multirow{2}{*}{\textbf{No.}}& \multicolumn{2}{c}{\textbf{Wavelength}} & \multirow{2}{*}{\textbf{No.}} & \multicolumn{2}{c}{\textbf{Wavelength}} \\
\cmidrule(lr){3-4}\cmidrule(lr){6-7}
 & & Range & Center & & Range & Center \\
\hline
B1 (Blue)   & B2  & 450--510   & 480  & B2  & 458--523   & 490  \\
B2 (Green)  & B3  & 530--590   & 560  & B3  & 543--578   & 560  \\
B3 (Red)    & B4  & 640--670   & 655  & B4  & 650--680   & 665  \\
B4 (NIR)    & B5  & 850--880   & 865  & B8  & 785--900   & 842  \\
B5 (SWIR-1) & B6  & 1570--1650 & 1610 & B11 & 1565--1655 & 1610 \\
B6 (SWIR-2) & B7  & 2110--2290 & 2200 & B12 & 2100--2280 & 2190 \\
\hline\hline
\end{tabular}
\end{center}
\end{table}

\subsection{Implementation Details}\label{sec:impl}
All experiments use the same co-registered inputs, include all connected date pairs, and adopt a fixed random seed. 
The invariant selection fraction is set to $\tau=0.4$ for both the initial and refined PIF populations, ensuring equal cardinalities and selection budgets. 
In addition, $\lambda=1$ and an ADMM penalty parameter of $c=0.1$ are used. 
The convergence tolerance is set to $10^{-8}$, with a maximum of 2000 iterations. 
Subsequently, the radiometric harmonization follows the monotone piecewise-linear formulation of \citet{DenaroLin2026Monotone}, using $K=3$ knots, $M=41$ quantile levels within $[0.01,0.99]$, $w_1=w_2=w_3=1$, and $w_4=0.5$. 
The harmonization procedure and all parameter settings remain identical across PIF selection methods to isolate the effect of CODE convex optimization.

Furthermore, six metrics are reported over the invariant population. 
Specifically, RMSE and PSNR quantify radiometric agreement between dates, SSIM assesses local structural agreement, and Pearson's correlation coefficient $\rho$ measures the linear association between corresponding per-band reflectance values. 
Then, two complementary spectral-angle measures distinguish temporal spectral alignment from distortion introduced by harmonization. 
Specifically, $\mathrm{SAM}_h$ measures the spectral angle between harmonized observations $\mathbf{X}'_j$ and $\mathbf{X}'_k$, assessing cross-date spectral alignment as the primary objective of harmonization. 
In contrast, $\mathrm{SAM}_{sc}$ \citep{denaro2026} measures the spectral angle between each original image and its corrected counterpart, quantifying the spectral distortion introduced by the transformation. 
All variants are evaluated using the same mask to ensure a consistent comparison.

\begin{figure*}[t]
\centering
\draftincludegraphics[width=1\linewidth]{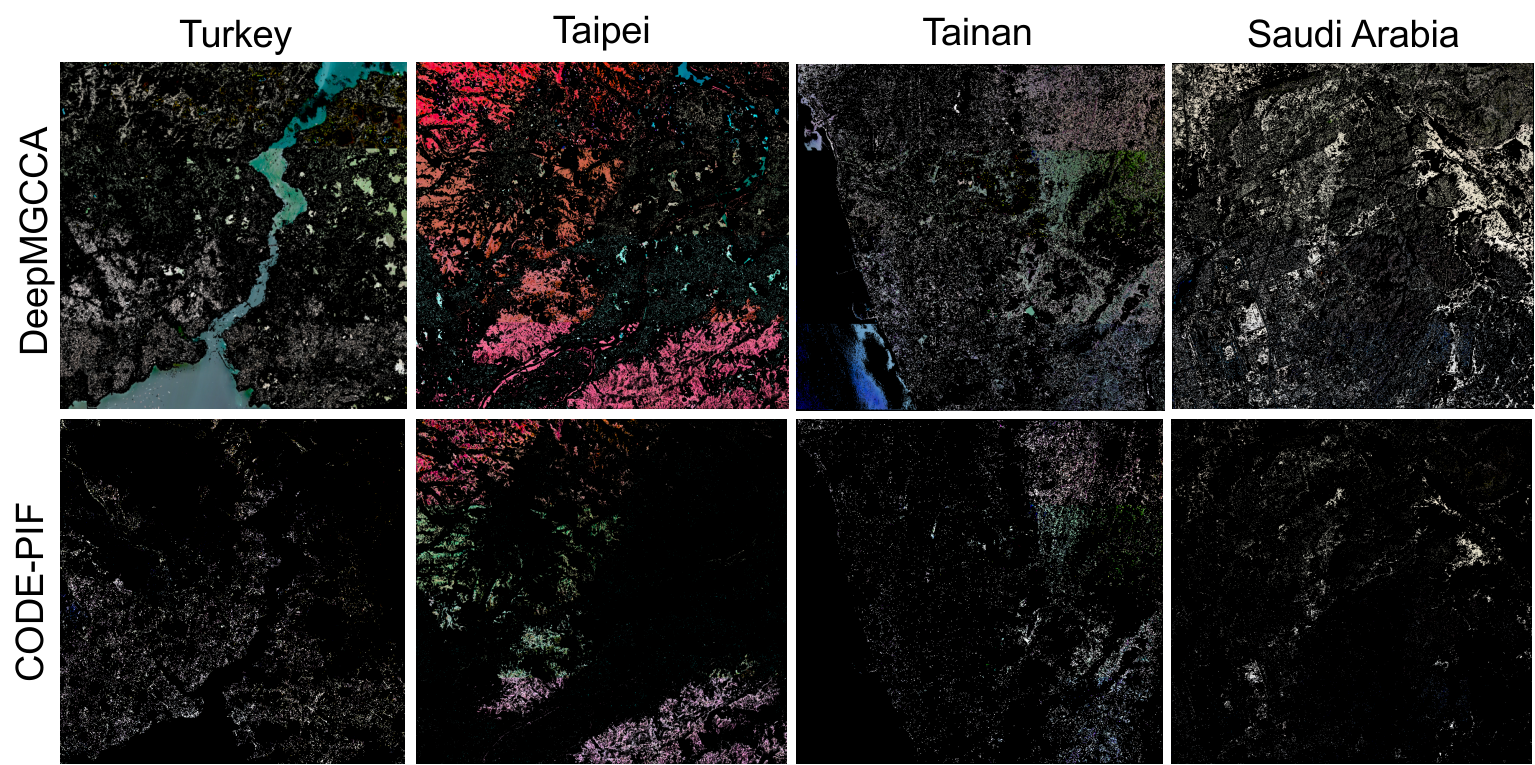}{1\linewidth}
\caption{PIF selection maps for Turkey, Taipei, Tainan, and Saudi Arabia, arranged from left to right. 
The top row shows the DeepMGCCA masks $\bm{\hat{x}}_{\text{DE}_j}$, while the bottom row shows the CODE-PIF selections obtained from the fused scores $\bs_j$. 
The selected pixels retain their colors from the corresponding reflectance composites, whereas unselected pixels are shown in black.}
\label{fig:pif}
\end{figure*}

\subsection{Qualitative Comparison of PIF Selection}\label{sec:pifs}

As presented in Fig.~\ref{fig:pif}, we compare the spatial distributions of the initial PIFs selected by DeepMGCCA (top row) and the refined PIFs obtained by CODE-PIF (bottom row) across Turkey, Taipei, Tainan, and Saudi Arabia. 
DeepMGCCA identifies candidate invariant pixels through representations shared across acquisition dates. 
However, consistency in the learned feature space does not necessarily imply agreement in the original reflectance values or spectral shapes, particularly in heterogeneous scenes containing built-up surfaces, vegetation, water, and bare soil.
The initial masks tend to comprise large, spatially contiguous regions. 
In Turkey, DeepMGCCA selects nearly all pixels corresponding to the Bosphorus Strait and the Sea of Marmara, while the Tainan mask similarly retains coastal waters along the western scene boundary. 
Although water typically exhibits low reflectance in the near-infrared and shortwave-infrared bands, its reflectance can vary with sun glint, turbidity, and surface conditions, potentially compromising its reliability as a radiometric reference. 
The Taipei mask is dominated by extensive vegetated mountain slopes in the northwest and southeast, along with scattered bodies of water. In Saudi Arabia, the initial selection covers broad areas of bright bare soil and urban development.

CODE-PIF substantially changes the spatial composition of the selected PIFs. 
Open-water regions initially selected by DeepMGCCA, including the Bosphorus and the Marmara coast in Turkey and the coastal waters in Tainan, are excluded from the refined masks and appear black in the bottom row. 
The retained pixels form spatially dispersed selections over built-up and bare surfaces rather than large contiguous regions. 
In Taipei, CODE-PIF preserves portions of the vegetated mountain slopes in the northwest and southeast while excluding most of the dense urban and riverine area. 
In Saudi Arabia, the refined mask retains a sparse subset of bare and urban surfaces, with concentrations toward the center-right of the scene.
These spatial changes are consistent with the two refinement mechanisms enabled by the fused score $\bs_j$. 
The initially selected PIFs exhibiting inconsistent multitemporal reflectance receive lower selection priority, whereas candidates excluded by DeepMGCCA can be included when their reflectance-derived discrepancy $\psi$ is sufficiently small. 
With the selection fraction held fixed, CODE-PIF adjusts the membership of the invariant population while preserving its prescribed size.

\begin{figure*}[t]
\centering
\draftincludegraphics[width=1\linewidth]{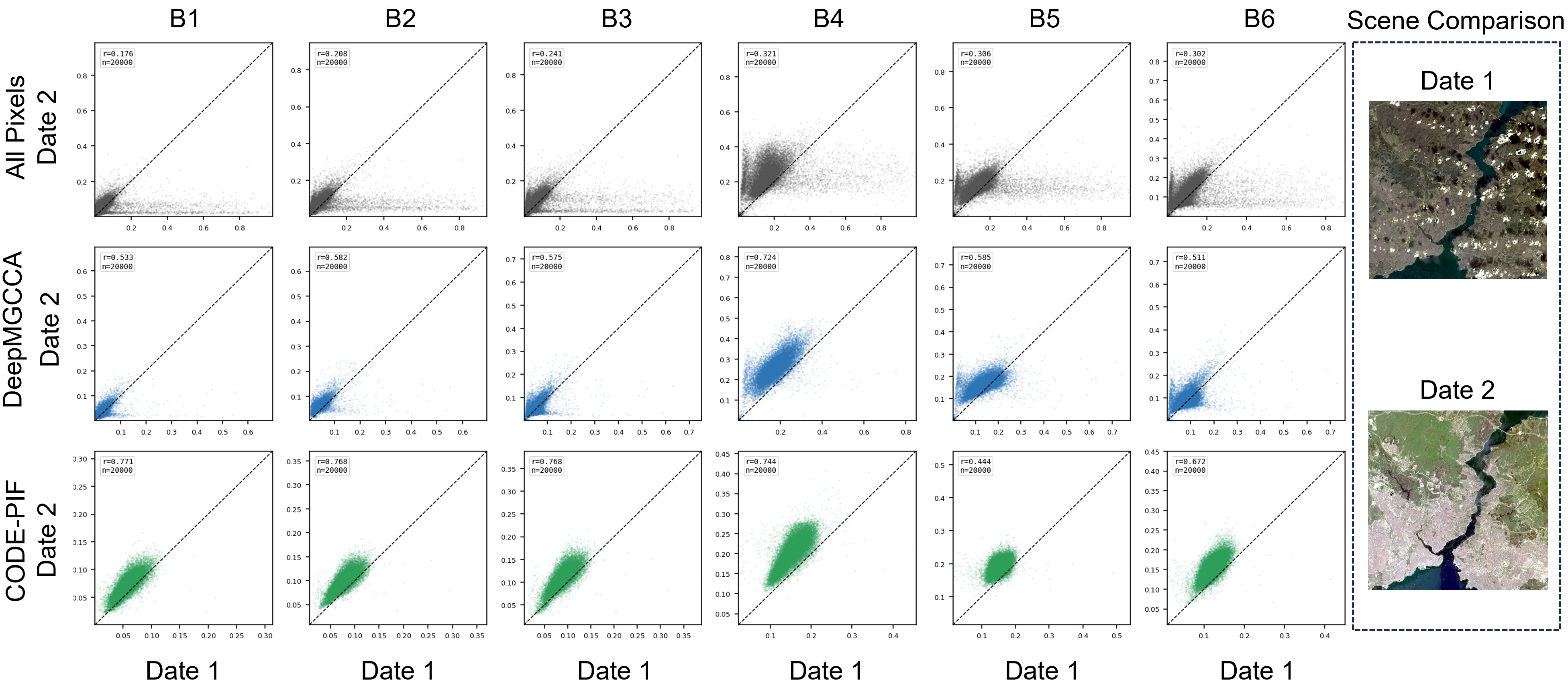}{1\linewidth}
\caption{Scatterplots comparing reflectance values between two representative dates in Turkey for bands B1--B6. 
The first row shows true-color images of Date~1 and Date~2. The second, third, and fourth rows show samples from all valid pixels, the DeepMGCCA PIFs, and the pixels selected as PIFs by CODE-PIF on both dates, respectively. 
Each scatterplot contains $n=20{,}000$ randomly sampled pixels and reports Pearson's $r$, with the 1:1 line shown as a dashed line. 
Points closer to the 1:1 line indicate better agreement in reflectance between the two dates.}
\label{fig:scatter_turkey}
\end{figure*}
\begin{figure*}[t]
\centering
\draftincludegraphics[width=1\linewidth]{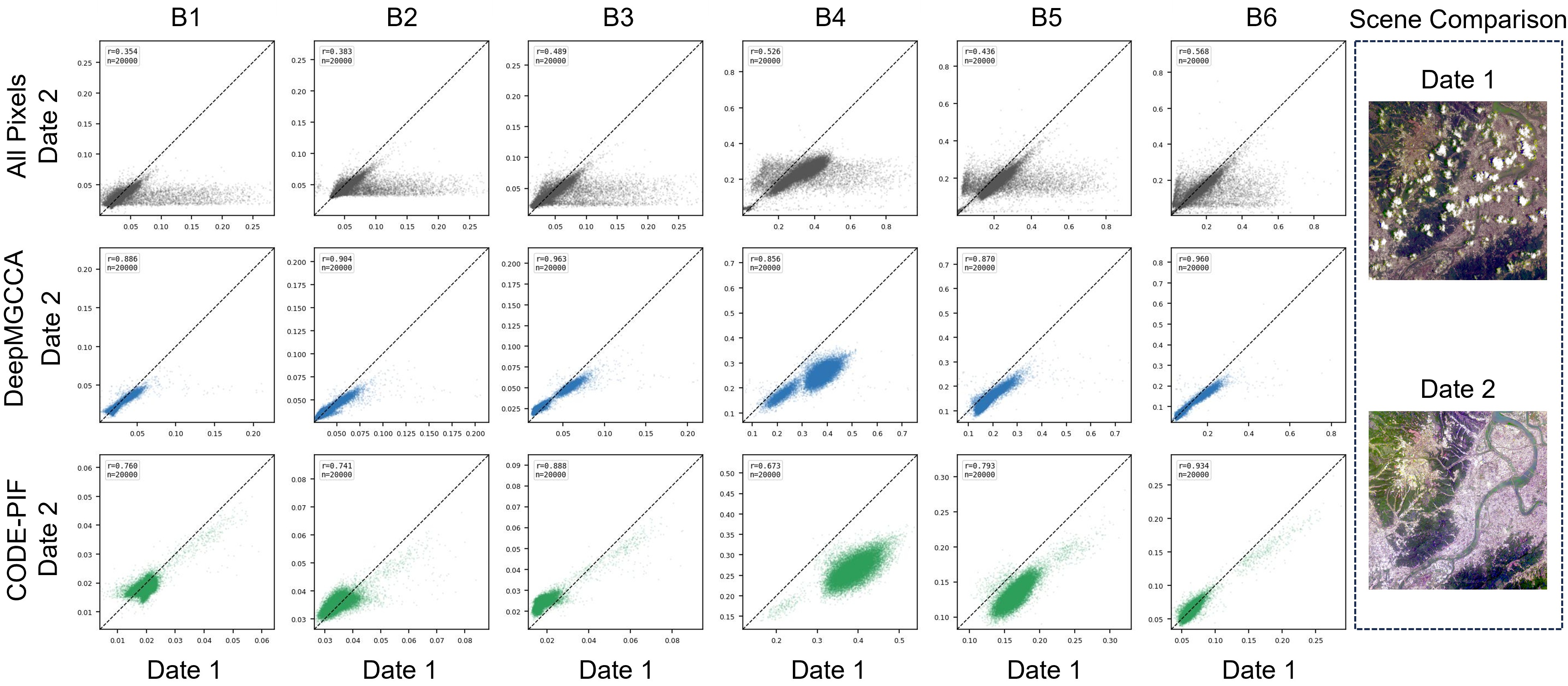}{1\linewidth}
\caption{Scatterplots comparing reflectance values between two representative dates in Taipei for bands B1--B6. The rows and symbols are the same as those in Fig.~\ref{fig:scatter_turkey}.}
\label{fig:scatter_taipei}
\end{figure*}

\begin{figure*}[t]
\centering
\draftincludegraphics[width=1\linewidth]{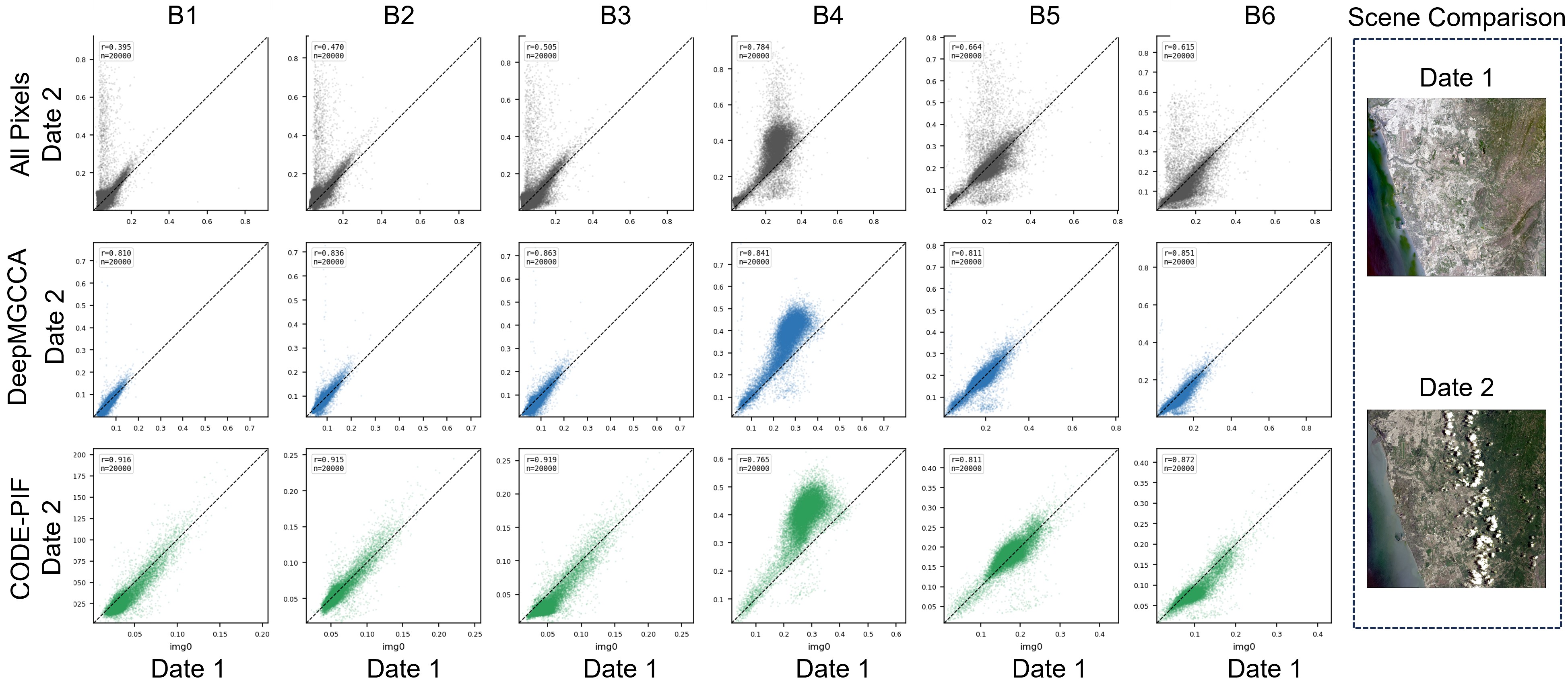}{1\linewidth}
\caption{Scatterplots comparing reflectance values between two representative dates in Tainan for bands B1--B6. The rows and symbols are the same as those in Fig.~\ref{fig:scatter_turkey}.}
\label{fig:scatter_tainan}
\end{figure*}

\begin{figure*}[t]
\centering
\draftincludegraphics[width=1\linewidth]{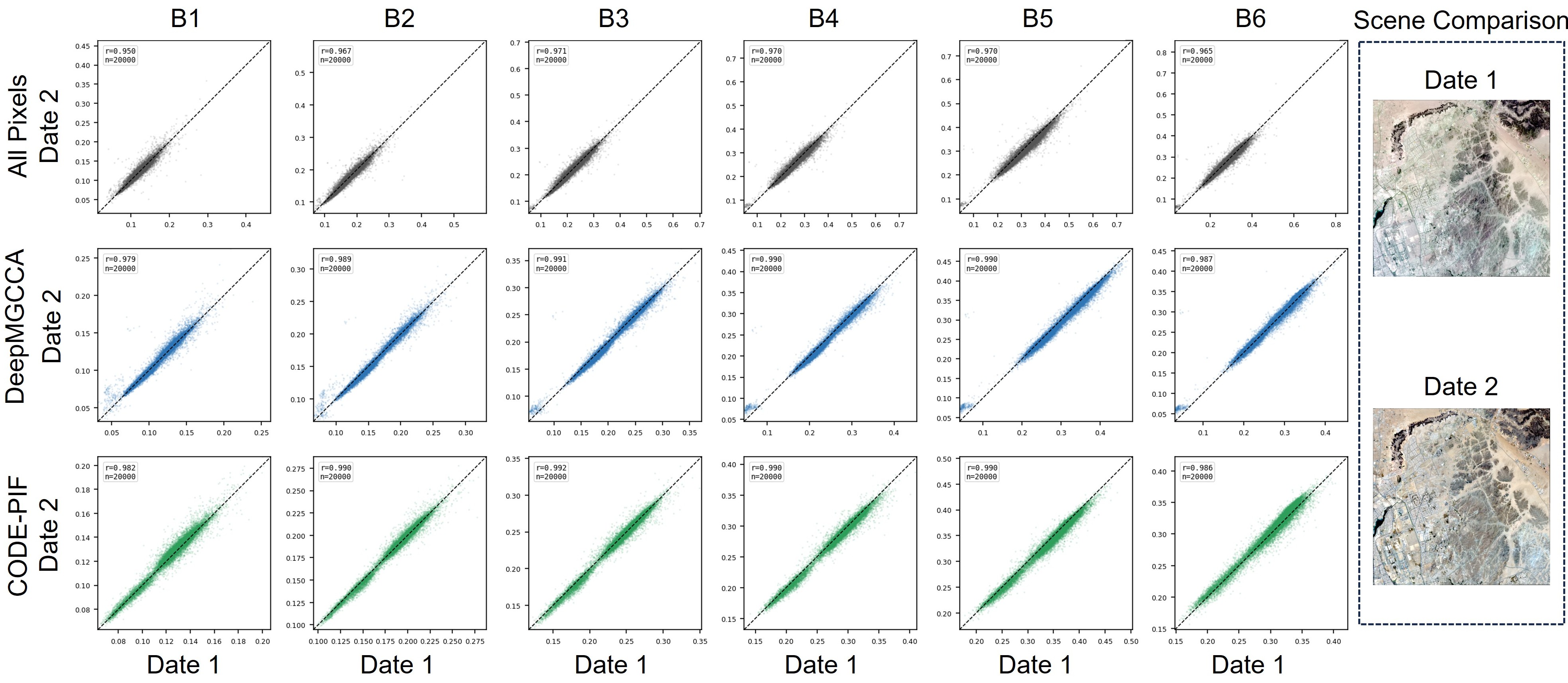}{1\linewidth}
\caption{Scatterplots comparing reflectance values between two representative dates in Saudi Arabia for bands B1--B6. The rows and symbols are the same as those in Fig.~\ref{fig:scatter_turkey}.}
\label{fig:scatter_saudi}
\end{figure*}

As shown in Figs.~\ref{fig:scatter_turkey}--\ref{fig:scatter_saudi}, we complement the spatial PIF maps with between-date reflectance scatterplots for one representative acquisition pair in each ROI. 
For each of the six bands, $n=20{,}000$ pixels are randomly sampled from each of three populations: all valid pixels, the DeepMGCCA PIFs, and the CODE-PIF PIFs. 
Equal sample sizes facilitate a controlled comparison of their scatter distributions. The full valid-pixel population contains both stable and changing surfaces, together with residual cloud and shadow contamination, resulting in substantial dispersion around the 1:1 line.
In the case of Turkey (cf. Fig.~\ref{fig:scatter_turkey}), the Date~1 contains scattered clouds and cloud shadows. 
The all-pixel distribution exhibits a pronounced horizontal tail corresponding to pixels with high reflectance on Date~1 but low reflectance on Date~2, yielding correlations of $r=0.176$--$0.321$. 
DeepMGCCA suppresses most of this tail and increases the correlations to $r=0.511$--$0.724$, although the selected samples remain broadly dispersed near the origin. 
CODE-PIF produces the most compact point distributions, with $r=0.744$--$0.771$ in B1--B4 and $r=0.672$ in B6. 
An exception occurs in SWIR-1 (B5), where the correlation decreases from 0.585 to 0.444. 
Despite its compact distribution, the CODE-PIF population spans a narrow reflectance range of approximately 0.1--0.2. 
Such a range restriction can reduce the Pearson correlation even when the between-date dispersion decreases. Across all bands, the CODE-PIF samples lie slightly above the 1:1 line, most noticeably in NIR (B4). 
This systematic displacement is consistent with inter-date gain and offset differences targeted by subsequent harmonization and does not, by itself, establish non-invariance.

On the other hand, for the Taipei scenario (cf. Fig.~\ref{fig:scatter_taipei}), the scattered cumulus clouds in Date~1 produce a horizontal tail in the all-pixel distribution, with correlations of $r=0.354$--$0.568$. 
The DeepMGCCA PIFs exhibit strong linear associations ($r=0.856$--$0.963$), partly reflecting their broad reflectance range across dark water, vegetation, and bright built-up surfaces. 
In addition, the two distinct clusters are particularly evident in the NIR band. CODE-PIF selects a narrower reflectance range, approximately 0.01--0.03 in B1, and yields lower correlations of $r=0.673$--$0.934$, with the largest reduction occurring in NIR. 
This pattern agrees with the lower aggregate Pearson correlation reported for CODE-PIF in Table~\ref{tab:quanti}. 
However, CODE-PIF achieves the lowest RMSE and $\mathrm{SAM}_h$ and the highest PSNR for this ROI. 
Therefore, Taipei demonstrates that reduced between-date reflectance discrepancies need not coincide with increased correlation when the selected population spans a narrower reflectance range.
The case in Tainan (cf. Fig.~\ref{fig:scatter_tainan}) demonstrates that a cloud band over the eastern mountains in Date~2 produces a vertical tail corresponding to pixels that are bright on Date~2 but not on Date~1. 
The all-pixel correlations range from $r=0.395$ to 0.784. DeepMGCCA removes most of this tail and increases the correlations to $r=0.810$--$0.863$, but retains isolated off-diagonal outliers in the visible bands and a low-reflectance tail below the main NIR distribution. 
CODE-PIF removes these outliers and increases the correlations in the visible bands (B1--B3: $r=0.915$--$0.919$) and SWIR-2 (B6: $r=0.872$), while SWIR-1 remains unchanged at $r=0.811$. 
In NIR, however, the refined distribution remains broad and displaced above the 1:1 line, with $r=0.765$ compared with 0.841 for DeepMGCCA. 
This residual variability is consistent with seasonal vegetation differences between acquisitions. 
It also agrees with the limited improvement reported for Tainan in Table~\ref{tab:quanti}, where RMSE decreases by only $0.5\%$, while SSIM and Pearson correlation decrease slightly.
Finally, for the Saudi Arabia case (cf. Fig.~\ref{fig:scatter_saudi}), the bare soil and built-up surfaces dominate this arid scene, and the all-pixel distributions already closely follow the 1:1 line ($r=0.950$--$0.971$). 
Therefore, both PIF populations exhibit nearly linear relationships ($r\geq0.979$), with correlation differences between DeepMGCCA and CODE-PIF not exceeding 0.003. 
The principal visual distinction concerns a small cluster of dark pixels that DeepMGCCA retains near the origin and below the main distribution in B1 and B4--B6. 
The CODE-PIF excludes these pixels, leaving a single elongated cluster along the 1:1 line. 
Thus, the corresponding improvements in Table~\ref{tab:quanti} are small but consistent across all metrics.

In conclusion, the scatterplots indicate that CODE-PIF suppresses cloud-related tails and isolated outliers remaining in the DeepMGCCA selections and generally produces more compact reflectance distributions. 
The bandwise correlations increase for most bands in Turkey, Tainan, and Saudi Arabia, but decrease in Taipei, Turkey B5, and Tainan B4. 
These exceptions should be interpreted alongside the restricted reflectance ranges and residual band-specific variability described above. 
Because the scatterplots represent a single acquisition pair and a random subsample, their correlation coefficients need not match the aggregate values in Table~\ref{tab:quanti}. 
Instead, they provide qualitative context for the quantitative trends discussed in Section~\ref{sec:performance}.

\begin{table*}[t]
\caption{Quantitative comparison of radiometric harmonization using PIFs selected by DeepMGCCA and CODE-PIF across the four ROIs. 
Results for the original observations (Orig.) are obtained from the DeepMGCCA evaluation files. Both PIF selection methods are evaluated using the same monotone piecewise-linear harmonization procedure \citep{DenaroLin2026Monotone}.
RMSE is expressed in reflectance units, PSNR in dB, and $\mathrm{SAM}_h$ and $\mathrm{SAM}_{sc}$ in degrees, while SSIM denotes spatial structural similarity evaluated using the PIF mask, and $\rho$ denotes Pearson correlation. 
Lower RMSE and $\mathrm{SAM}_h$ and higher PSNR, SSIM, and $\rho$ indicate better performance. $\mathrm{SAM}_{sc}$ quantifies the spectral change introduced by harmonization and is reported descriptively. 
The best values of the performance metrics within each ROI are highlighted in bold.}
\label{tab:quanti}
\centering
\renewcommand{\arraystretch}{1.2}
\setlength{\tabcolsep}{2.2mm}
\small
\begin{tabular}{c l c c c c c c}
\hline\hline
ROI & Variant
& RMSE$\downarrow$
& PSNR$\uparrow$
& SSIM$\uparrow$
& SAM$_h$$\downarrow$
& $\rho\uparrow$
& SAM$_{sc}$ \\
\hline

\multirow{3}{*}{\emph{Turkey}}
 & Orig.
 & 0.0322 & 31.12 & 0.8179
 & 6.668 & 0.6714 & --- \\
 & DeepMGCCA
 & 0.0243 & 33.37 & 0.8474
 & 5.474 & 0.6677 & 4.165 \\
 & CODE-PIF
 & \textbf{0.0180} & \textbf{35.26}
 & \textbf{0.8707}
 & \textbf{3.695} & \textbf{0.6967}
 & \textbf{1.360} \\
\hline

\multirow{3}{*}{\emph{Taipei}}
 & Orig.
 & 0.0366 & 31.07 & 0.8769
 & 6.020 & \textbf{0.8668} & --- \\
 & DeepMGCCA
 & 0.0172 & 37.05 & 0.9248
 & 2.950 & 0.8095 & 3.643 \\
 & CODE-PIF
 & \textbf{0.0138} & \textbf{40.09}
 & \textbf{0.9357}
 & \textbf{2.262} & 0.5835
 & \textbf{2.149} \\
\hline

\multirow{3}{*}{\emph{Tainan}}
 & Orig.
 & 0.0335 & 31.34 & 0.8431
 & 9.453 & \textbf{0.8831} & --- \\
 & DeepMGCCA
 & 0.0215 & 34.92 & \textbf{0.8747}
 & 5.182 & 0.8585 & 7.980 \\
 & CODE-PIF
 & \textbf{0.0214} & \textbf{35.26}
 & 0.8712
 & \textbf{4.621} & 0.8035
 & \textbf{4.498} \\
\hline

\multirow{3}{*}{\emph{Saudi Arabia}}
 & Orig.
 & 0.0239 & 34.21 & 0.9090
 & 3.050 & 0.9482 & --- \\
 & DeepMGCCA
 & 0.0136 & 38.29 & 0.9236
 & 1.455 & 0.9474 & 1.672 \\
 & CODE-PIF
 & \textbf{0.0123} & \textbf{39.14}
 & \textbf{0.9511}
 & \textbf{1.315} & \textbf{0.9491}
 & \textbf{1.541} \\
\hline\hline
\end{tabular}
\end{table*}

\subsection{Quantitative Comparison of PIF Selection}
\label{sec:performance}

As demonstrated in Table~\ref{tab:quanti}, we summarize the quantitative results for the original observations and the PIF populations selected by DeepMGCCA and CODE-PIF across the four ROIs. 
Both selection methods use the same invariant fraction, ensuring a shared selection budget for assessing the contribution of convex refinement. 
The proposed CODE-PIF achieves lower RMSE and between-date spectral angle ($\mathrm{SAM}_h$), together with higher PSNR, than DeepMGCCA in all four regions. 
Averaged across the ROIs, RMSE decreases from 0.0192 to 0.0164, corresponding to an approximately $14.5\%$ reduction, while $\mathrm{SAM}_h$ decreases from $3.77^\circ$ to $2.97^\circ$, an approximately $21.0\%$ reduction. 
The mean PSNR increases from 35.91 to 37.44~dB.
The regional improvements are consistent with the scatterplots, with the most pronounced gains observed in Turkey and Taipei.
Specifically, Turkey exhibits the largest relative reductions in RMSE ($25.9\%$) and $\mathrm{SAM}_h$ ($32.5\%$), consistent with the exclusion of water pixels from the initial selection in Fig.~\ref{fig:pif} and the more compact reflectance distributions in Fig.~\ref{fig:scatter_turkey}.
Similarly, Taipei shows substantial improvement, achieving a $19.8\%$ RMSE reduction and the largest PSNR gain of 3.04~dB.
The gains are more modest in Saudi Arabia, where the initial DeepMGCCA population already exhibits strong radiometric consistency (Fig.~\ref{fig:scatter_saudi}), leaving less room for improvement and yielding a $9.6\%$ RMSE reduction.
In Tainan, the improvement is further limited to a $0.5\%$ RMSE reduction and a 0.34~dB PSNR gain, consistent with the residual NIR variability that remains after refinement (Fig.~\ref{fig:scatter_tainan}).
This progression from substantial to marginal gains suggests that the effectiveness of convex refinement depends on both the quality of the initial PIF population and the radiometric characteristics of the scene.

The consistent improvements in radiometric and spectral agreement are accompanied by more variable responses in SSIM and Pearson correlation.
For SSIM, CODE-PIF improves performance in Turkey, Taipei, and Saudi Arabia, while Tainan exhibits a slight decrease from 0.8747 to 0.8712.
Pearson correlation similarly increases in Turkey and Saudi Arabia but decreases from 0.8095 to 0.5835 in Taipei and from 0.8585 to 0.8035 in Tainan.
These correlation reductions are consistent with the restricted reflectance ranges and residual band-specific variability illustrated in Figs.~\ref{fig:scatter_taipei} and~\ref{fig:scatter_tainan}.
Thus, despite smaller between-date errors, the refined populations do not necessarily exhibit stronger linear correlation or greater local structural similarity.
A separate consideration is the spectral change introduced by harmonization, as measured by $\mathrm{SAM}_{sc}$.
CODE-PIF yields lower $\mathrm{SAM}_{sc}$ values than DeepMGCCA in all four ROIs, indicating smaller changes in spectral shape between the original and corrected images.
This result complements the improvements in between-date agreement by characterizing the spectral modification introduced by the transformation.
However, because $\mathrm{SAM}_{sc}$ compares each image with its corrected counterpart, it does not independently establish the intrinsic stability of the selected PIFs.
Therefore, RMSE, PSNR, and $\mathrm{SAM}_h$ remain the primary evidence of improved consistency across dates.

In conclusion, the spatial maps and scatterplots provide qualitative context for these quantitative findings by illustrating how convex refinement changes the DeepMGCCA selection, including the exclusion of affected water and cloud-contaminated pixels.
Also, these observations support the effectiveness of CODE-PIF in improving radiometric and spectral agreement, while also revealing scene-dependent trade-offs in individual metrics.
Therefore, PIF selection should be assessed jointly in terms of reflectance agreement, spectral stability, spatial structure, and linear correlation.

\subsection{Discussion}\label{sec:discussion}

The reliability of CODE-PIF depends on the quality of both the initial DeepMGCCA selection and the reflectance-derived change scores. 
The pixelwise data-fitting map does not explicitly model local spatial relationships and may overlook variations in reflectance magnitude that preserve spectral shape. 
Furthermore, its temporal reference and correlation weights may also become unreliable under widespread surface change, persistent contamination, or substantial registration errors. 
These conditions can limit the ability of convex refinement to correct initial selection errors, particularly when both information sources provide misleading indications of invariance.
The experimental protocol also defines the scope of the reported improvements. 
Using a fixed invariant fraction enables a controlled comparison of candidate re-ranking and membership replacement, but does not determine the optimal size of the invariant population. 
Furthermore, radiometric and spectral performance is assessed using a common monotone piecewise-linear harmonizer \citep{DenaroLin2026Monotone}. 
The results therefore quantify the contribution of PIF refinement within this calibration pipeline and do not independently establish the physical invariance of every selected pixel. 
Independent temporal verification and evaluation, along with additional harmonization procedures, would provide further evidence of the reliability and broader applicability of CODE-PIF.

\section{Conclusions and Future Works}\label{sec:conclusion}

In this article, we propose CODE-PIF (cf. Fig. \ref{fig:general}), an unsupervised CODE framework that refines DeepMGCCA-derived PIFs through mathematically simple convex optimization for multitemporal surface reflectance harmonization. 
Despite the capability of DeepMGCCA in nonlinear feature extraction across different dates, spectrally distinct observations may exhibit similar representations in its learned feature space, potentially limiting the reliability of PIF selection. 
However, the DeepMGCCA-derived PIFs preserve useful structural information that can serve as a prior for subsequent refinement.
Therefore, the proposed CODE-PIF first uses DeepMGCCA to obtain a rough PIF selection, which is then enhanced in the convex optimization stage. 
During the convex optimization stage (cf. Section \ref{sec:code}), we formulate the data-fitting term by combining reflectance discrepancies, correlation-weighted spectral angles, and temporal-reference change metrics to promote radiometric and spectral consistency of the selected PIFs across acquisition dates (cf. Section \ref{sec:data_fitting}).
To promote the desired structural properties, we employ a mathematically simple $\bQ$-quadratic-norm regularizer \citep{CODE} that extracts useful information from the coarse DeepMGCCA-derived PIFs and incorporates it into the final PIFs.  
The extensive experiments across four geographically distinct regions demonstrate improved average radiometric and spectral agreement compared with DeepMGCCA, thereby achieving state-of-the-art performance. 
The average RMSE decreased from 0.0192 to 0.0164, PSNR increased from 35.91 to 37.44 dB, and the between-date spectral angle decreased from 3.77° to 2.97°.
In the future, we will focus on developing an interpretable architecture for multitemporal surface reflectance harmonization, facilitating explicit analysis of how model updates affect radiometric and spectral consistency. 
Interpretability is also important for downstream applications, including land-cover and blue carbon ecosystem monitoring \citep{CODEHCD,CODEMM}, where spurious spatial or spectral variations may be misinterpreted as genuine surface changes.
Motivated by recent advances in deep unfolding for remote sensing \citep{COS2A}, we plan to investigate a convex formulation for multitemporal surface reflectance harmonization and unfold the associated iterative solver into a trainable network.
This architecture will combine the mathematical interpretability of the underlying optimization model with the flexibility of deep learning to improve PIF selection.

\section*{Acknowledgements}
This study was supported by the Emerging Young Scholar Program (the 2030 Cross-Generation Young Scholars Program) of National Science and Technology Council (NSTC), Taiwan, under Grant NSTC 115-2628-E-006-001. We thank the National Center for Theoretical Sciences (NCTS) and the National Center for High-performance Computing (NCHC) for providing the computing resources.
\bibliographystyle{cas-model2-isprs-first}

\bibliography{ref}

\end{document}